\documentclass[12pt]{article}
\usepackage{lingmacros}
\usepackage[english]{babel}
\usepackage{graphicx}

\large \baselineskip = 20 true pt

\begin{document}
\begin{center}
{\Large \bf User Authorization in a System with a Role-Based Access Control on the Basis of 
the Analytic Hierarchy Process} \vspace{0.5cm}
\end{center}

\begin{center}
S.V. Belim, S.Yu. Belim, N.F. Bogachenko, A.N. Kabanov\\
Dostoevsky Omsk State University, Omsk, Russia

 \vspace{0.5cm}
\end{center}

\begin{center}
{\bf Abstract}
\end{center}

The problem of optimal authorization of a user in a system with a role-based access control 
policy is considered. The main criterion is to minimize the risks of permission leakage. 
The choice of the role for authorization is based on the analytic hierarchy process. 
The substantiation of a choice of criteria for formation of a hierarchy of the first level 
is given. An algorithm for calculating weight coefficients is presented, based on the 
quantitative characteristics of the role graph and not dependent on subjective expert 
evaluations. The complexity is estimated and the scalability of the proposed algorithm 
is discussed.

\section{Introduction}

The basic idea of the role-based access control policy is that rights and privileges 
(we will call them permissions) are granted to the user not directly, but through assigning 
a certain role to the user. Together with the role, the user receives a certain set 
of permissions assigned to the given role. Each session of the user in the system requires 
his authorization with one or more allowed roles for him.

In this paper we consider the problem of selecting one or more roles in a situation where the 
list of permissions that a user must have is known, and a given set of permissions can be granted
by several roles. The indicated task arises both at the stage of granting potentially possible
roles to the user, and when authorizing in each specific session of work. In both cases, we will
call the posed problem the authorization problem.

When designing a role-based access control, the system administrator must decide on the optimal
organization for the requested access. Given that the security of information in computer systems
is made up of confidentiality, accessibility and integrity, the following selection criteria for
roles can be formulated:

1. Give the user as much permissions as possible - make the system as open as possible.

2. Give the user a minimum of permissions - make the system as closed as possible.

In connection with the mutual inconsistency of these criteria, the choice of the optimal version 
of granting roles to the user is a rather complex problem. In \cite{b1} it is proposed 
to consider the linear combination of the functions of "openness" (efficiency of information
access) $F_1$ and "closedness" (the effectiveness of information confidentiality) $F_2$ as an
objective function of the problem:
\[
F = c_1F_1 + c_2F_2.
\]
But with this approach, there is a significant difficulty in determining the quantitative
characteristics of the weight coefficients $ñ_1$ and $ñ_2$, and the functions $F_1$ 
and $F_2$ themselves. In addition, the proposed criteria need to be expanded and detailed. 
In particular, such indicators as integrity and accessibility of information are not involved. 
In terms of the main channels of information leakage that can arise as a result of authorization,
the criterion of "closedness" should be decomposed into components. In this situation, linear
programming models for decision making become unacceptable. But the existence of several criteria
and alternative solutions makes it possible to suppose the applicability of the analytic 
hierarchy process \cite{b2,b3,b4,b5,b6,b7}. At the same time, it is important to determine the quantitative indicators for constructing a numerical scale of preferences for possible alternatives (in our case, roles that have the required permissions for the user).

\section{Statement of the problem and criteria of the hierarchy}

Let a role-based access control policy with hierarchical organization of multiple roles is
implemented in the computer system. This means that on a set of roles the submission/domination
ratio is given that generates a hierarchy of roles. The permissions between roles 
are distributed according to this hierarchy.

A new user u appears in the system which requires a set PU of permissions to work. It is required
to authorize this user for some role. There are two possible solutions. The first is to authorize
the user for one of the existing roles. The second solution is to create a new role, specifically
for this user. The second approach, of course, provides a higher level of security, but leads 
to a number of problems. First, it can lead to uncontrolled growth of many roles. Secondly, it
requires solving additional tasks to determine the hierarchy of roles. Third, it can eliminate 
all the advantages of the role-based access control model if for each role there will be only one
user authorized to it. In addition, allowing the modification of the hierarchical structure 
of multiple roles requires additional analysis of hidden channels of information leakage 
\cite{b8, b9}. The purpose of the proposed approach is to develop an algorithm that allows 
new users to be entered without changing the role structure of the system.

Let's imagine the authorization problem in terms of the analytic hierarchy process. 
In the simplest case, it will be a task with one hierarchical level: the solution, the criteria
for the hierarchy of level 1, and alternatives. Expected result is ranking of roles by degree 
of preference from the point of view of system administration. The hierarchy top or solution 
is the choice of a specific role for user authorization. Alternatives are the roles 
$r_1, r_2, ..., r_k$, which include in their list of permissions the required set $PU$ of
permissions.

Let us pass to the selection of criteria. At the initial stage of the model building, 
two opposite factors can be chosen: "openness" and "closedness" of the system. But, as noted
earlier, this approach is too generalized and complex for quantitative estimates. Another option 
is to formulate criteria based on an analysis of possible channels of information leakage, 
leaving issues of integrity and accessibility beyond the scope of the review, focusing on
confidentiality of information. We formulate a minimum risk requirement for solving this 
problem: the authorization of a new user should be carried out in such a way as to minimize 
the risks of information leakage. We indicate the main channels of information leakage that 
can arise as a result of authorization:

1. Obtaining additional permissions attributed to roles, but not intended for the user.

2. User authorization for additional roles subordinate to this role.

We will compare roles on the specified factors: "additional permissions" and "subordinate roles".
The advantage of this approach is that for both criteria there are quantitative estimates
determined by a given hierarchy of roles - this is the power of the set of additional permissions
provided to the user when authorizing for this role, and the power of the set of roles over which
this role is dominant. The smaller these values, the more preferable this role is from the
standpoint of confidentiality. Then pairwise comparison matrices of the level of alternatives can be obtained numerically, without constructing a scale of preferences and attracting experts.

\section{Filling the pairwise comparison matrix}

Let user u need to get a set $PU$ of permissions. We write out all the roles containing those
permissions: $r_1, r_2, ...,r_k$. At this stage of the problem analysis, we will compare the 
roles based on the previously mentioned criteria:

1. A is the set of additional permissions granted to the user when authorizing for this role.
Quantitative estimate is $|A|$ (the power of the set $A$): the smaller $|A|$, the higher 
the level of confidentiality.

\begin{figure}[ht]
\centering
\includegraphics[width=0.3\textwidth]{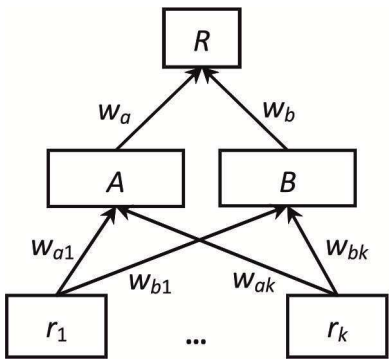}
\caption{The hierarchy of the level 1, constructed according to two criteria}
\label{fig1}
\end{figure}

2. $B$ is the set of roles over which this role dominates. Quantitative estimate is $|B|$ 
(the power of the set $B$): the smaller $|B|$, the higher the level of confidentiality.
The decision tree is shown in Fig.\ref{fig1}.

According to the analytic hierarchy process \cite{b4}, if the relative weight coefficients 
$w_a$, $w_b$, $w_{a1}, ..., w_{ak}$, $w_{b1}, ..., w_{bk}$ ($i = 1, ..., k$) are known, 
then the $i$-th role should be chosen with probability equal to $i$-th combined weight factor:
\[
P(r_i) = w_a \times w_{ai} + w_b \times w_{bi}.
\]
To minimize the risks of information leakage, the user $u$ must be authorized for the role $r_i$
with the maximum value of $P(r_i)$. Once again, we emphasize that the calculation of the weight
coefficients is carried out from the point of view of the system administrator.

To determine the relative weight coefficients, the pairwise comparison method \cite{b4} is used.
For the top of the hierarchy (the solution $R$) and the criteria (in our case, these are the
criteria $A$ and $B$) the pairwise comparison matrices are compiled. The dimension $k$ of the
current pairwise comparison matrix $M$ is determined by the number of factors being compared 
(for the solution this is the number of criteria, for each criterion this is the number of roles). The diagonal elements $[M]_{ii}$ of the matrix $M$ are equated to 1, and the elements 
$[M]_{ij}$ show how many times the factor $i$ is more significant than the factor $j$. 
Then the columns of the pairwise comparison matrix are normalized, and the elements of the new normalized matrix $M^*$ are calculated by the formula:
\[
[M^*]_{ij}=\frac{[M]_{ij}}{[M]_{1j}+...+[M]_{kj}}.
\]
Finally, the weight of each factor $w_i$ is calculated as the arithmetic mean of 
the corresponding row in the normalized matrix $M^*$:
\[
w_i=\frac{1}{k}\left([M^*]_{i1}+...+ [M^*]_{ik}\right)
\]

We consider first the pairwise comparison matrix $M_r$ for the top $R$ and the factors $A$ 
and $B$. We assume that the additional permissions obtained directly (factor $A$) create 
the danger of information leakage $s$ times more often than through the subordinate roles 
(factor $B$). This means that the factor $A$ in comparison with factor $B$ is s times more
preferable or more significant from the point of view of the attacker. Since our task is 
to obtain weights that reflect the administrator's preferences, this value must be inverted, 
which means that the elements of the pairwise comparison matrix are $[M_r]_{ab} = 1/s$. 
As a result, the pairwise comparison matrix Mr has the form:
\[
M_r=\left(\begin{array}{cc}
            1 & 1/s \\
            s & 1 
          \end{array}
\right)
\]
Let's move on to the next level of the hierarchy. The quantitative estimate for criterion $A$ 
is the value inverse to the power of the set of additional permissions provided to the user 
when authorizing for the selected role. Then the elements of the pairwise comparison matrix 
$M_a$ for determining the coefficients $w_{a1}, ...,w_{ak}$ will be found by the formula:
\[
[M_a]_{ij}=\frac{1/dp_i}{1/dp_j}=\frac{dp_j}{dp_i},
\]
here $dp_i$ is the number of permissions of role $r_i$, excluding the required set $PU$ 
of permissions. Note that if for some $i$ the equality $dp_i = 0$ holds, then user $u$ 
is authorized for the role $r_i$ precisely, and this completes the calculations. 
Similarly, the quantitative estimate for criterion B is the value inverse to the power of 
the set of roles over which the selected role is dominant. To determine the coefficients 
$w_{b1}, ...,w_{bk}$, we construct the pairwise comparison matrix $M_b$, in which
\[
[M_b]_{ij}=\frac{1/dr_i}{1/dr_j}=\frac{dr_j}{dr_i},
\]
here $dr_i$ is the number of roles over which the role $r_i$ is dominant. To avoid zero values 
and, as a consequence, infinitely large values in weight coefficients, we assume that each role
dominates itself, that is, the minimum value of the power of the set of subordinate roles will 
be equal to one. Thus, we assume that the dominance relation is reflexive (non-strict).

\section{Consistency o the pairwise comparison matrices}

It is known that in the case when the pairwise comparison matrix is ideally consistent: 
$[M]_{ij} = [M]_{is} \times [M]_{sj}$ (for any $i$, $j$, $k$), the columns of the corresponding
normalized matrix are obtained the same \cite{b4}. Then the weight coefficients are 
$w_i = [M^*]_{ij}$ for any $j$, and for their calculation it is sufficient to normalize only one
column of the pairwise comparison matrix.

The pairwise comparison matrices $M_r$, $M_a$, $M_b$ were filled by us without involving 
the mechanism of expert evaluations, and their ideal consistency follows in an obvious manner 
from the filling algorithm itself. Indeed, based on the hierarchy of roles and the distribution 
of permissions, we attributed to the compared alternatives (roles) some numerical characteristics
that are convenient to represent in vector form:
\[
P=\left(\begin{array}{c}
            1/dp_i \\
            \vdots\\
            1/dp_k
          \end{array}
\right), \ \ \
R=\left(\begin{array}{c}
            1/dr_i \\
            \vdots\\
            1/dr_k
          \end{array}
\right).
\]
Then, the elements of the matrices $M_a$ and $M_b$ were calculated as the ratios of 
the corresponding coordinates of these vectors. Obviously, the following equalities hold:
\[
[M_a]_{is} \times [M_a]_{sj}
=\frac{dp_s}{dp_i}\times\frac{dp_j}{dp_s}=\frac{dp_j}{dp_i}=[M_a]_{ij},
\]
\[
[M_b]_{is} \times [M_b]_{sj}
=\frac{dr_s}{dr_i}\times\frac{dr_j}{dr_s}=\frac{dr_j}{dr_i}=[M_b]_{ij}.
\]
The ideal consistency of the matrix $M_r$ follows from the fact that the two-dimensional
reciprocal matrix is always ideally consistent. But in this case the vectors of the weight
coefficients $W_a$ and $W_b$ are none other than the normalized vectors $P$ and $R$. Indeed:
\[
w_{ai}=\frac{[M_a]_{ij}}{[M_a]_{1j}+...+[M_a]_{kj}}
=\frac{dp_j/dp_i}{dp_j/dp_1+...+dp_j/dp_k}
=\frac{1/dp_i}{1/dp_1+...+1/dp_k},
\]
\[
W_a=\left(\begin{array}{c}
            w_{a1} \\
            \vdots\\
            w_{ak}
          \end{array}
\right)=\frac{P}{||P||}=
=\left(\begin{array}{c}
            \frac{1/dp_1}{1/dp_1+...+1/dp_k} \\
            \vdots\\
            \frac{1/dp_k}{1/dp_1+...+1/dp_k}
          \end{array}
\right).
\]
\[
w_{bi}=\frac{[M_b]_{ij}}{[M_b]_{1j}+...+[M_b]_{kj}}
=\frac{dr_j/dr_i}{dr_j/dr_1+...+dr_j/dr_k}
=\frac{1/dr_i}{1/dr_1+...+1/dr_k},
\]
\[
W_b=\left(\begin{array}{c}
            w_{b1} \\
            \vdots\\
            w_{bk}
          \end{array}
\right)=\frac{R}{||R||}=
=\left(\begin{array}{c}
            \frac{1/dr_1}{1/dr_1+...+1/dr_k} \\
            \vdots\\
            \frac{1/dr_k}{1/dr_1+...+1/dr_k}
          \end{array}
\right).
\]
The vector of weight coefficients of the level 1 is obtained by normalizing the first column of 
the matrix $M_r$:
\[
W=\left(\begin{array}{c}
            w_{a} \\
            w_{b}
          \end{array}
\right)=
\left(\begin{array}{c}
            1/(1+s) \\
            s/(1+s)
          \end{array}
\right).
\]

It should be noted that the main difficulty of the analytic hierarchy process lies precisely 
in the search for each level of the hierarchy of positive weight coefficients 
$w_1, ..., w_k$ satisfying the normalization condition: $w_1 + ... + w_k = 1$. 
The formulas used to calculate the weights are based on the following fact: the weight vector 
is the normalized eigenvector of an ideally consistent pairwise comparison matrix corresponding 
to its maximal eigenvalue equal to the dimension of the matrix $k$ \cite{b4}. 
An ideal consistency of the pairwise comparison matrix is possible, for example, 
if experts fill only the first row of the matrix, and the remaining elements are calculated 
based on the relation: $[M]_{ij} = [M]_{1j} / [M]_{1i}$ \cite{b3}. In practice, with the 
classical "element-by-element" formation of pairwise comparison matrices by experts, 
one cannot count on its consistency. Nevertheless, the weight vector in the analytic 
hierarchy process is also equal in this case to the eigenvector corresponding to 
the maximum eigenvalue, which for the inconsistent matrix is already strictly greater than $k$. 
But the equality used is heuristic, which means that the application of the analytic hierarchy
process in the case of an inconsistent pairwise comparison matrix contains a "model" error in
computing the weight vector. To evaluate it, the author of the method proposes to use a special
numerical index - the "consistency index" \cite{b4}. Unfortunately, the value of this index 
allows us to judge the magnitude of the resulting "model" error only indirectly \cite{b3}.

In this approach, to solving the authorization problem, there is no need to form pairwise
comparison matrices, and the analytic hierarchy process, in addition to reducing the complexity, 
is deprived of the main source of criticism, a "model" error that arises from the inconsistency 
of expert judgments.

\section{Algorithm of selecting the role for user authorization}

We formally write down the algorithm for solving the authorization problem. As before, 
we assume that the additional permissions received by the user directly create the danger 
of information leakage $s$ times more often than through authorization for subordinate roles. 
Let the set of permissions which is necessary for user $u$ in the current session is set by 
the set $PU$. The permissions of role r are accessible through the field (set) $r.p$, and the 
list of subordinate roles is accessible through the field (set) $r.r$. The algorithm steps can 
be written in the following form.

1. Create a set $RU$ of roles that are available to user $u$ for authorization, according to the rule: if $PU\subseteq r_i.p$ , then $r_i\in RU$ .

2. Each role $r_i\in RU$ is put in correspondence with the value $dp_i$ equal to the number of additional permissions of the role $r_i: dp_i=|r_i.p|-|PU|$. If there is a role for which $dp = 0$, then user $u$ must be authorized for this role, and the algorithm's work ends.

3. Each role $r_i\in RU$ is put in correspondence with the value $dr_i$ equal to the number of roles over which the role $r_i$ is dominant: $dr_i=|r_i.r|$ (providing $r_i\in r_i.r$ ).

4. Calculate the coordinates of the weight vectors of the level of alternatives 
(vectors $W_a$ and $W_b$):
\[
w_{ai}=\frac{1/dp_i}{1/dp_1+...+1/dp_k},\ \ 
w_{bi}=\frac{1/dr_i}{1/dr_1+...+1/dr_k},
\]
(here $k=|RU|$).

5. Calculate the coordinates of the weight vector of the level of criteria (vector $W$):
\[
w_a=\frac{1}{1+s},\ \ w_b=\frac{s}{1+s}.
\]
6. Calculate for each role $r_i\in RU$ the probability of its selection by the system administrator to authorize the user $u$:
\[
P(r_i)=w_a\times w_{ai}+w_b\times w_{bi}.
\]

7. Find the role $r_{max}\in RU$ such that
\[
P(r_{max})=\max_{r_i\in RU} P(r_i).
\]

8. Authorize user $u$ for the role $r_max$.

We estimate the algorithmic complexity of the presented algorithm.

{\bf Statement.} The complexity of the algorithm for solving the authorization problem based 
on the analytic hierarchy process is $O(max(n \times m, n^2))$, here $n$ is the number of roles, 
$m$ is the number of permissions.

{\bf Proof.} We denote the complexity of the $i$-th step of the algorithm by $T_i$ 
($i = 1, ..., 8$), and the complexity of the preparatory stage by $T_0$. Then the total 
complexity of the algorithm $T$ is calculated by the formula: $T = T_0 + T_1 + ... +T8$. 
Notice that $|PU|\leq m$, $|RU|\leq n$.

For calculations, it is necessary to compute the powers of the sets $r_i.p$ and $r_i.r$ 
for each $r_i$ ($i = 1, ...,n$) in the system. We denote the complexity of these operations 
by $T_p$ and $T_r$. Obviously, regardless of the implementation of the data structure that 
defines the hierarchy of roles, $T_p = O(n \times m)$ and $T_r = O(n^2)$. The complexity 
of calculating the power of the set  $PU$ also doesn't cause difficulties: $T_{PU} = O(m)$. 
Thus, $T_0 = O(n \times m) + O(n2) + O(m) = O(max(n \times m, n^2))$.

We denote the complexity of solving the problem of checking the inclusion of one set in another 
($PU\subseteq r_i.p$ ) by $T_\subseteq$ . Then $T1 = nT_\subseteq$, $T_2 = O(n)$, $T_3 = O(n)$, 
$T_4 = O(n)$, $T_5 = O(const)$, $T_6 = O(n)$, $T_7 = O(n)$, $T_8 = O(const)$.

The complexity $T_\subseteq$  depends on the way the set is implemented. Obviously, 
$T_\subseteq\geq O(m)$. We show that we can choose a way to form the set, which allows 
us to obtain equality for the implementation of the operation "is contained". Indeed, 
the universal set for the sets $PU$ and $r_i.p$ is the set of all permissions. If the set 
is given by a binary vector (array) of dimension $m$, in which the $j$-th coordinate is 1, 
if the permission $p_j$ belongs to the set, and 0 otherwise, then the "is contained" operation 
is realized as follows: $PU\subseteq r_i.p$ if and only if 
($[PU]_i = 0) \vee (([PU]_i = 1) \wedge ([r_i.p]_i = 1))$. The complexity of this test is $O(m)$. Therefore, $T_\subseteq = O(m)$ and $T_1 = O(n \times m)$.

As a result, we get: $T = O(max(n \times m, n^2)) + O(n \times m) + O(n) + O(n) + O(n) + O(const) + O(n) + O(n) + O(const) = O(max(n \times m, n^2))$. Q.E.D.

{\bf Note.} It should be noted that the main complexity of the algorithm lies in the preparatory
stage: $T_0 = O(max(n \times m, n2))$. If the powers of the sets $r_i.p$, $r_i.r$ and $PU$ 
are known in advance, then $T = O(n \times m)$. In this case, again, the complexity is determined
by the decision of the choice of roles that are "suitable" for authorizing the user $u$. 
In the case when this question is solved in advance, the complexity of the algorithm becomes
linear: $T = O(n)$.

\section{Scalability}

In the approach presented, two criteria are used to estimate the confidentiality of information. Obviously, the method is scalable and can easily be extended to any number of criteria.

We add to the proposed set of criteria factors to estimate the accessibility and integrity 
of information. It can be a "total set of permissions" that includes the total number 
of permissions of the current role. The more the power of such a set, the more preferable 
this role is in terms of accessibility. As for integrity, to take into account this component 
of information security, it will be necessary to introduce the criterion of the same name 
and estimates the roles, rather attributed to them sets of permissions, pairwise for the 
integrity of information. If we try, as for the three previous criteria, to obtain 
a quantitative estimate, then we can single out a set of permissions that are potentially 
dangerous in terms of integrity. Their total count for the current role will give the desired
numerical value. The lower this indicator, the more preferable the role.

In addition to the criteria generated by the requirements of information security, one can 
consider the factor characterizing the costs of management. The criterion of "manager costs" 
is determined by the cost of the role $r_i$ in managing subordinate roles: 
$(dm_i + 1)^\alpha\lambda^\alpha$, here $\alpha$ and $\lambda$ are some constants, 
$dm_i$ is the number of roles directly subordinate to the role $r_i$ \cite{b10}. 
Obviously, the role for which this value is minimal is preferable.

Here it should be noted again that for the presented criteria there are quantitative estimates
determined by a given dominance relation on the set of roles and allowing to fill the pairwise
comparison matrices automatically.

In determining the weights of the criteria level, there is some subjectivity since, even with 
the requirement of the ideal consistency of the matrix $M_r$, it is necessary to determine 
the degree of preference of the first criterion in comparison with the others (to form 
the first row of the matrix). These quantities are configurable parameters of the algorithm. 
Their variation and recalculation of probabilities can give the system administrator additional
information about the stability of the formed hierarchy of many roles against changes 
in external conditions.

\section{Conclusion}

As can be seen from the proposed methodology, the analytic hierarchy process allows to
significantly automate the process of selecting a role for user authorization in order 
to obtain the necessary permissions. However, the recommendations of the method should 
not be taken as a final solution. It is more reasonable to arrange a list of suitable roles 
by not increasing the probability of choice. Then the decision should be made by the system
administrator, based on the meaning of the permissions themselves and additional restrictions 
on the ownership of role or permission that may exist in the system. Thus, the proposed approach 
is a decision support system.

It should also be noted that the use of the analytic hierarchy process proposed in the article 
is based only on the properties of the system itself, without external subjective expert 
estimates. Such use is new, not previously used. And this approach allows us to run the analysis 
of the system in an automatic mode, not only when authorizing a new user in the system, but also 
to check the security of the system as a whole.

The estimation of the method complexity carried out in the article allows to speak about 
the absence of difficulties in the program implementation of this approach and high speed.

In conclusion, we note that the analysis of hierarchical structures existing in the information
system allows obtaining additional information on the distribution of access rights and requires
the development of mathematical models of security taking into account the hierarchy of their
elements \cite{b8,b9,b11}.


\end{document}